\documentclass[aps,prc,
twocolumn,
showpacs,groupedaddress]{revtex4}
\usepackage{graphicx}
\usepackage{amsmath}
\usepackage{amsfonts}
\usepackage{amssymb}
\begin{document}
\title{Covariant calculation of mesonic baryon decays}
\author{T. Melde}
\author{W. Plessas}
\author{R. F. Wagenbrunn}
\affiliation{%
Theoretical Physics, Institute for Physics, University of Graz,
Universit\"atsplatz 5, A-8010 Graz, Austria}
\date{\today}
\begin{abstract}
We present covariant predictions for $\pi$ and $\eta$ decay modes of
$N$ and $\Delta$ resonances from relativistic constituent-quark models
based on one-gluon-exchange and Goldstone-boson-exchange dynamics. The
results are calculated within the point-form approach to
Poincar\'e-invariant relativistic quantum mechanics applying a 
spectator-model decay operator. The direct predictions of the
constituent-quark models for covariant $\pi$ and $\eta$ decay widths show
a behaviour completely different from previous ones calculated in
nonrelativistic or so-called semirelativistic approaches. 
It is found that the present theoretical results agree with 
experiment only in a few cases but otherwise always remain 
smaller than the experimental data (as compiled by the Particle Data 
Group). Possible reasons for this behaviour are discussed with regard 
to the quality of both the quark-model wave functions and the 
mesonic decay operator.
\end{abstract}
\maketitle

\section{Introduction}
\label{intro}

Hadronic transitions between baryon states represent a wide field of 
physical phenomena to be understood ultimately on the basis of quantum 
chromodynamics (QCD), the fundamental theory of strong interactions. 
While there is a wealth of experimental data available, theory lags 
behind with regard to a comprehensive explanation. This is mainly 
due to the persisting difficulties of solving QCD rigorously in the
low- and intermediate-energy regimes. There one has to resort to
effective theories or models based as far as
possible on the genuine properties of QCD. Furthermore, they should
provide a comprehensive framework covering also other hadron
phenomena (e.g., interactions with electroweak probes etc.). 
The quark-model description of light and strange bary\-ons has seen a 
number of interesting and important new developments over the last few years.
In addition to the traditional constituent-quark model (CQM), whose hyperfine
interaction derives from one-gluon exchange (OGE)~\cite{Capstick:1986bm},  
alternative types of CQMs have been suggested such as the ones based on 
instanton-induced (II) 
forces~\cite{Loring:2001kx,Loring:2001ky} or Goldstone-boson-exchange 
(GBE) dynamics~\cite{Glozman:1996fu}. The GBE 
CQM~\cite{Glozman:1998ag,Glozman:1998fs} aims at incorporating
the basic properties of low-energy QCD, as following
from the spontaneous breaking of chiral symmetry (SB$\chi$S). 

Properties of baryon resonances should be calculated in a 
fully relativistic approach. In this paper, the theory is formulated along 
relativistic, i.e. Poincar\'e-invariant, quantum 
mechanics~\cite{Keister:1991sb}. Specifically,
we adhere to its point-form version~\cite{Dirac:1949,Leutwyler:1977vy}, since this allows 
to calculate observables in a manifestly covariant 
manner~\cite{Klink:1998pr}. This approach is a-priori distinct from a 
field-theoretic treatment. It relies on a relativistically invariant mass operator 
with the interactions included according to the Bakamjian-Thomas 
construction~\cite{Bakamjian:1953}. In this way all 
the required symmetries of special relativity can be fulfilled.
Relativistic CQMs have already been applied
in the description of electroweak nucleon 
form factors~\cite{Cardarelli:1995dc,Wagenbrunn:2000es,Glozman:2001zc,%
Boffi:2001zb,Merten:2002nz,Julia-Diaz:2003gq} and
electric radii as well as magnetic moments of all octet and decuplet 
baryon ground states~\cite{Berger:2004yi,Metsch:2004qk}. In this context
the point-form approach has turned out surprisingly successful.
Here we are interested if an analogous treatment of strong decays also
leads to a satisfactory description of this type of hadronic reactions,
in agreement with existing experimental data.

Mesonic resonance decays have always been considered as a big challenge,  
with early attempts dating back to the
sixties~\cite{Becchi:1966yt,Mitra:1967yt,Faiman:1968js,Faiman:1969at,%
Feynman:1971wr}. With the refinement of CQMs over the years more  
studies on different aspects of mesonic decays have been performed. 
In the course of the past two decades a number of valuable insights 
have thus been gained by various groups, e.g., in 
Refs.~\cite{Koniuk:1980vy,Kumano:1988ga,LeYaouanc:1988aa,%
Stancu:1988gb,Stancu:1989iu,Capstick:1993th,Capstick:1994kb,%
Geiger:1994kr,Ackleh:1996yt}. 
In the focus of interest have been, notably, the performance of various
CQMs as well as the adequacy of different decay operators for the
mechanism of meson creation/emission. Despite the considerable
efforts invested one has still not yet arrived at a satisfactory 
microscopic explanation especially of the $N$ and $\Delta$ resonance
decays. Also complementary attempts outside the CQM approach have not
succeeded much better with hadron decays, and 
more generally, with providing a comprehensive working model
of low-energy hadronic physics based on QCD. This 
situation is rather unsatisfactory from the theoretical side, 
especially in view of the large amount of experimental data 
accumulated over the past years and the ongoing high-quality 
measurements at such facilities as JLAB, MAMI and others (for a an 
overview of the modern developments see the proceedings of the
recent $N^*$ 
Workshops~\cite{Drechsel:2001,Dytman:2003pi,Bocquet:2004}). 

Up till now, specifically the GBE CQM has already been put to some tests in 
calculating mesonic decays of resonances of light and strange 
baryons in a semi-relativistic
framework~\cite{Krassnigg:1999ky,Plessas:1999nb,Theussl:2000sj}. 
These studies have revealed that relativistic effects have 
a big influence on the results, both in an elementary-emission and 
a quark-pair-creation model of the decay operator. In the present work,
we now perform 
a covariant calculation of $\pi$ and $\eta$ decay modes of $N$ and 
$\Delta$ resonances. At this instance, we use a rather simplified 
model for the decay operator. Our primary goal has been to set up 
a fully relativistic (covariant) CQM formulation of mesonic decays;
later on one may still improve on the decay operator. In particular,
we assume a decay operator in the point-form spectator
model (PFSM) with a pseudo-vector coupling. It has been seen
in previous studies that such an operator includes effective
many-body contributions due to the symmetry requirements of Poincar\'e
invariance (especially in order to satisfy translational invariance
of the transition amplitude)~\cite{Melde:2004qu,Melde:2004qa}. We 
produce the corresponding predictions for decay widths from the GBE 
CQM and analogous results from a CQM with a OGE hyperfine interaction,
namely, the relativistic version of the Bhaduri-Cohler-Nogami
CQM~\cite{Bhaduri:1980fd} as
parametrized in ref.~\cite{Theussl:2000sj}. In addition a comparison 
is provided with results from the II CQM obtained with a similar
(spectator-model) decay operator in a Bethe-Salpeter 
approach~\cite{Metsch:2004qk}. The relativistic results are also 
contrasted to several nonrelativistic and so-called semirelativistic 
calculations. Partially, preliminary results have already been presented
in proceedings contributions~\cite{Melde:2002ga,Melde:2004xj,Melde:2004ce}.

In Sect.~\ref{sec:theory} we outline the theory for a 
covariant calculation of the mesonic decay widths from a 
relativistic CQM. In Sects.~\ref{sec:results} and~\ref{sec:comparison}
we present the results of our calculation and discuss their qualitative
and quantitative features with a comparison to decay 
calculations along other models and/or approaches.
In the Appendix some details of the relativistic
point-form calculation of mesonic baryon decays are given. 

\section{Theory}{\label{sec:theory}}

Generally, the decay width $\Gamma$ of a particle is defined by
\begin{equation}
\label{eq:decwidth}
	\Gamma=2\pi \rho_{f}\left| F\left(i\rightarrow
	f\right)\right|^{2},
\end{equation}
where $ F\left(i\rightarrow f\right)$ is the transition amplitude
and $\rho_{f}$ is the phase-space factor. In order to get the total 
decay width one has to average over the initial and to sum over
the final spin-isospin projections. 

In nonrelativistic calculations the strong decays of hadron resonances
are treated with a transition amplitude that is not Lorentz-invariant.
Consequently, one is left with an arbitrary choice of the phase-space 
factor~\cite{Kumano:1988ga,Geiger:1994kr,Kokoski:1987is}. In the rest
frame of the decaying resonance, either a purely nonrelativistic
form,
\begin{equation}
\label{eq:nrphase}
\rho_{f}=\frac{M' m}{M}q,
\end{equation}
or the relativistic form,
\begin{equation}
\label{eq:relphase}
\rho_{f}=\frac{E' \omega_{\rm m}}{M}q,
\end{equation}
has been used. In Eq. (\ref{eq:nrphase}), $M$ is the mass of the initial
resonance while $M'$ and $m$ are the masses of the final state 
and the emitted meson, respectively; q is the magnitude of the 
momentum transfer. Correspondingly, in Eq. (\ref{eq:relphase}),
$E'$ and $\omega_{\rm m}$ are the energies of the decay products.
An alternative choice was made by Capstick and Roberts~\cite{Capstick:1993th} 
using the phase-space factor
\begin{equation}
\label{eq:kiphase}
\rho_{f}=\frac{{\tilde M'}{\tilde m}}{{\tilde M}}q,
\end{equation}
first introduced by Kokoski and Isgur~\cite{Kokoski:1987is} for meson
decays. Here the quantities with tilde represent some effective
(parametrized) masses. Clearly, the particular choice of the phase-space
factor has a pronounced effect on the final results. The ambiguity 
concerning the phase-space factor is immediately resolved by imposing 
relativistic invariance on the formalism. This can evidently be done either
along a relativistic field theory or in relativistic (Poincar\'e-invariant)
quantum mechanics.

In the present work we formulate a Poincar\'e-invariant 
description of the decay amplitude. Out of the possible forms of 
relativistic dynamics minimally affected by 
interactions~\cite{Dirac:1949,Leutwyler:1977vy}
we make use of the point form. In this case one has the 
advantage that only the four-momentum operator $\hat P^{\mu}$ 
contains interactions. Consequently, the generators of the
Lorentz transformations 
remain purely kinematic and the theory is manifestly 
covariant~\cite{Klink:1998pr}. The interactions are introduced into 
the (invariant) mass operator following the Bakamjian-Thomas
construction~\cite{Bakamjian:1953}. Hereby the free mass operator
${\hat M}_{\rm free}$ is replaced by a full mass operator ${\hat M}$
containing an interacting term ${\hat M}_{\rm int}$:
\begin{equation}
\label{eq:masses}
\hat M_{\rm free} \rightarrow \hat M = \hat M_{\rm free} + \hat M_{\rm int}.
\end{equation}
The four-momentum operator is then defined by multiplying the mass 
operator ${\hat M}$ by the four-velocity operator $\hat V^{\mu}$
\begin{equation}
    \hat P^{\mu}={\hat M}\hat V^{\mu} \, .
\end{equation}
In the point form, following the Bakamjian-Thomas construction, the 
four-velocity operator is kinematic and thus remains independent
of interactions, i.e. $\hat V^{\mu}=\hat V^{\mu}_{\rm free}$.
The eigenstates of the four-momentum operator
$\hat P^{\mu}$ are simultaneous eigenstates also of the mass
operator ${\hat M}$ and the four-velocity operator $\hat V^{\mu}$; 
this is simply a consequence of Poincar\'e-invariance.
For a given baryon state of mass $M$ and
total angular momentum $J$ with z-projection $\Sigma$ the
eigenvalue problem of the mass operator reads
\begin{equation}
    {\hat M}\left|V,M,J,\Sigma\right>
    =M\left|V,M,J,\Sigma\right>.
\end{equation}
Here we have written the eigenstates in obvious notation as
$\left|V,M,J,\Sigma\right>$, where $V$ indicates the four eigenvalues
of $V^{\mu}$, of which only three are independent. 
Alternatively we can express these eigenstates also as
\begin{equation}
\left|V,M,J,\Sigma\right> \equiv \left|P,J,\Sigma\right>,
\label{eq:eigenstates}
\end{equation}
where $P$ represents the four eigenvalues of $\hat P^{\mu}$, whose 
square gives the invariant mass operator.

For the actual calculation in the point form it is advantageous to
introduce a specific basis of free three-body states, the so-called
velocity states, by
\begin{eqnarray}
&&\left|v;\vec{k}_1,\vec{k}_2,\vec{k}_3;\mu_1,\mu_2,\mu_3\right\rangle
=U_{B(v)}
\left|k_1,k_2,k_3;\mu_1,\mu_2,\mu_3\right\rangle
\nonumber \\
&&=\sum_{\sigma_1,\sigma_2,\sigma_3}
\prod\limits_{i=1}^3D^{\frac{1}{2}}_{\sigma_i\mu_i}[R_W(k_i,B(v))]
\left|p_1,p_2,p_3;\sigma_1,\sigma_2,\sigma_3\right\rangle .
\nonumber \\
& &
\label{eq:velstates}
\end{eqnarray}
Here $B\left(v\right)$, with unitary representation 
$U_{B\left(v\right)}$,
is a boost with four-velocity $v$ on the free three-body states
$\left|k_1,k_2,k_3;\mu_1,\mu_2,\mu_3\right\rangle$ in the 
centre-of-momentum system, i.e., for which $\sum{\vec k_{i}}=0$.
The second line in Eq.~(\ref{eq:velstates}) expresses the 
corresponding Lorentz transformation as 
acting on general three-body states 
$\left|p_1,p_2,p_3;\sigma_1,\sigma_2,\sigma_3\right\rangle$. 
The momenta $p_{i}$ and $k_{i}$ are related by 
$p_{i}=B\left(v\right)k_{i}$, where
$k_{i}=\left(\omega_{i},{\vec k}_{i}\right)$.
The $D^{\frac{1}{2}}$ are the spin-$\frac{1}{2}$ representation 
matrices
of Wigner rotations $R_{W}\left(k_{i},B\left(v\right)\right)$.
It is advantageous to use the velocity-state basis (instead of the 
basis of general free three-body states) since in this case Lorentz 
transformations rotate all particles by the same angle 
and the spin coupling can be done in the usual way 
\cite{Keister:1991sb,Klink:1998hc}. Some further details concerning
velocity states are given in the Appendix.  

The relativistic transition amplitude for the mesonic decay of a baryon
resonance $\left|V,M,J,\Sigma\right>$ to the nucleon ground state 
$\left|V',M',J',\Sigma'\right>$ is defined by the reduced matrix element 
of the mesonic decay operator $\hat D_m$

\begin{widetext}%
\begin{eqnarray}
F\left(i\rightarrow
	f\right)&=&\left<V',M',J',\Sigma'\right|{\hat D}_{m,\rm rd}
\left|V,M,J,\Sigma\right>
\nonumber \\
&=&
\frac{2}{MM'}\sum_{\sigma_i\sigma'_i}\sum_{\mu_i\mu'_i}{
\int{
d^3{\vec k}_2d^3{\vec k}_3d^3{\vec k}'_2d^3{\vec k}'_3
}}
\sqrt{\frac{\left(\sum \omega_i\right)^3}
{2\omega_1 2\omega_2 2\omega_3}}
\sqrt{\frac{\left(\sum \omega'_i\right)^3}
{2\omega'_1 2\omega'_2 2\omega'_3}}
\nonumber\\
&&
{
\Psi^\star_{M'J'\Sigma'}\left({\vec k}'_i;\mu'_i\right)
\prod_{\sigma'_i}{D_{\sigma'_i\mu'_i}^{\star \frac{1}{2}}
\left\{R_W\left[k'_i;B\left(V'\right)\right]\right\}
}}
\nonumber \\
&&
\left<p'_1,p'_2,p'_3;\sigma'_1,\sigma'_2,\sigma'_3\right|{\hat 
D}_{m,\rm rd}
\left|p_1,p_2,p_3;\sigma_1,\sigma_2,\sigma_3\right>
\nonumber \\
&&
\prod_{\sigma_i}{D_{\sigma_i\mu_i}^{\frac{1}{2}}
\left\{R_W\left[k_i;B\left(V\right)\right]\right\}
}
\Psi_{MJ\Sigma}\left({\vec k}_i;\mu_i\right)
\, .
\label{transampl}
\end{eqnarray}
\end{widetext}
The wave functions $\Psi^{\star}_{M'J'\Sigma'}$ and
$\Psi_{MJ\Sigma}$ denote
velocity-state representations of the baryon states 
$\left<P',J',\Sigma'\right|$ and $\left|P,J,\Sigma\right>$, 
respectively. For the decay operator $\hat D_{m,\rm rd}$ we
assume a spectator model with pseudovector coupling and express its 
matrix element by
\begin{multline}
\left<p'_1,p'_2,p'_3;\sigma'_1,\sigma'_2,\sigma'_3\right|
{\hat D}_{m,\rm rd}\left|p_1,p_2,p_3;\sigma_1,\sigma_2,\sigma_3\right>
=\\
3 \frac{i g_{qqm}}{2m_1\left(2\pi\right)^{\frac{3}{2}}}
 \sqrt{\frac{M^{3}M'^{3}}{
    \left(\sum{\omega_{i}}\right)^{3}
    \left(\sum{\omega'_{i}}\right)^{3}}}\\
    {\bar u}\left(p'_1,\sigma'_1\right)
\gamma_5\gamma^\mu \lambda_m
u\left(p_1,\sigma_1\right)
Q_\mu
\\
2p_{20}\delta\left({\vec p}_2-{\vec p}'_2\right)
2p_{30}\delta\left({\vec p}_3-{\vec p}'_3\right)
 \delta_{\sigma_{2}\sigma'_{2}}
   \delta_{\sigma_{3}\sigma'_{3}}
\label{eq:hadrcurr}\, ,
\end{multline}
where $g_{qqm}$ is the meson-quark coupling constant, $\lambda_m$ 
the flavour operator for the particular decay channel,
and $m_1$ the mass of the quark coupling to the generated meson. In the
actual calculations for the theoretical predictions of the CQMs to be
presented in the next Section the meson-quark coupling constant is assumed
to be $g_{qqm}^2/{4 \pi}=0.67$. This value is consistent with the one used
in the parametrization of the GBE CQM (for both the $\pi$-quark as well as
the $\eta$-quark couplings)~\cite{Glozman:1998ag}. The dependence of the
results for strong decay widths on the size of the meson-quark coupling
constant is discussed in Sect.~\ref{sec:comparison}.

Eq.~(\ref{eq:hadrcurr}) defines the spectator-model decay operator,
here specifically in point form (PFSM).
The transition amplitude in Eq.~(\ref{transampl}) is Poincar\'e-invariant, 
and the overall momentum conservation $P_{\mu}-P_{\mu}'=Q_{\mu}$ has already
been exploited in the integral. Regarding the spectator model of the decay 
operator it should be noted that in PFSM the impulse delivered
to the quark that emits the meson is not equal to the impulse
delivered to the nucleon as a whole. However, the momentum transfer 
$\tilde q$ to this single quark is uniquely determined from the momentum $Q$
transferred to the nucleon and the two spectator 
conditions (cf. Eq.~(\ref{eq:spectcond}))~\cite{Melde:2004qu}.
The square-root normalization factor has been introduced in accordance with the 
previous PFSM studies in the electromagnetic 
case~\cite{Wagenbrunn:2000es,Glozman:2001zc,Boffi:2001zb,Berger:2004yi}.

\section{Direct predictions for the decay widths}{\label{sec:results}}

\begin{figure*}
\includegraphics{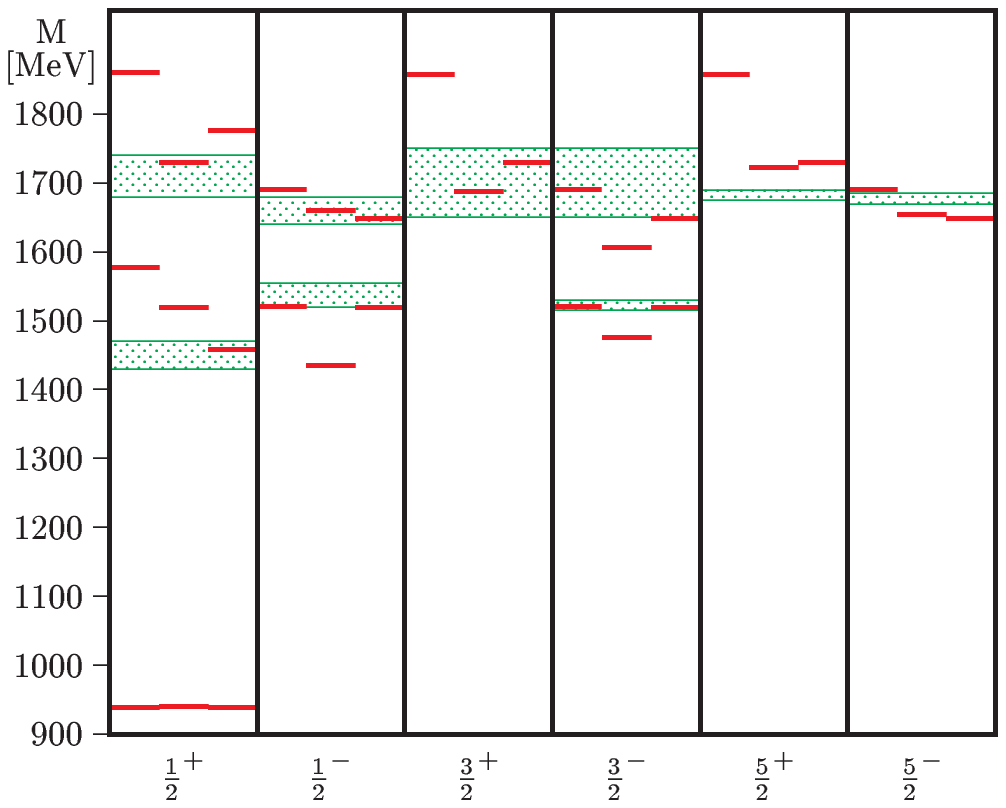}
\includegraphics{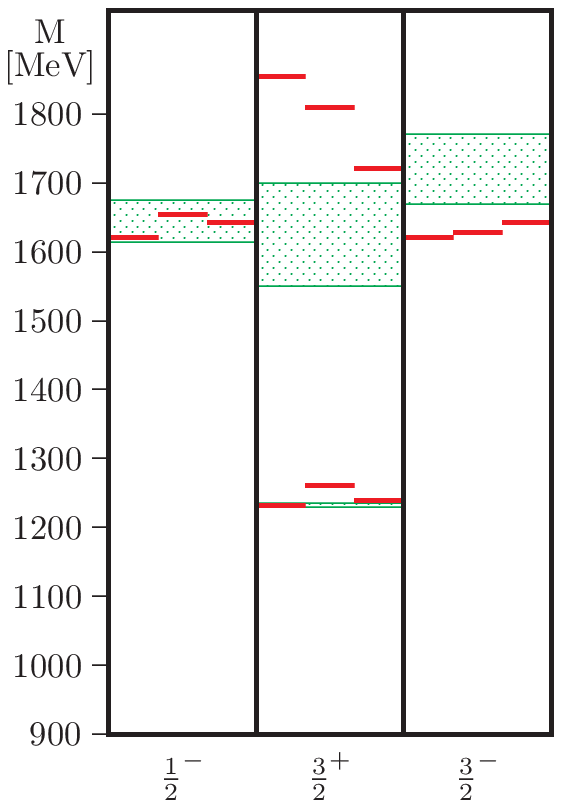}
\caption{\label{fig:spectrum}
$N$ (left panel) and $\Delta$ (right panel) spectra from three different
types of relativistic CQMs. In each column the left horizontal lines represent
the results of the relativistic version of the Bhaduri-Cohler-Nogami
CQM~\protect\cite{Theussl:2000sj}, the middle ones of the II CQM
(Version A)~\protect\cite{Loring:2001kx}, and the right ones of the GBE
CQM~\protect\cite{Glozman:1998ag}. The shadowed boxes give the experimental
data with their uncertainties after the latest compilation of the
PDG~\protect\cite{Eidelman:2004wy}.
}
\end{figure*}

It is well known that the underlying quark dynamics of CQMs has
a pronounced effect on the baryon spectra~\cite{Plessas:2003av}. In 
Fig.~\ref{fig:spectrum} we show a comparison of the $N$ and $\Delta$ spectra
for three different CQMs. While the $N$-$\Delta$ splittings are 
correct in all cases, it is seen that only the GBE CQM succeeds 
simultaneously to reproduce the proper level ordering of positive- and 
negative-parity excitations. Now, it is interesting to learn how the CQMs 
with different dynamics predict the widths of various mesonic decay 
modes.

In Table~\ref{tab1} we present the covariant predictions of the relativistic
CQMs for $\pi$ decay widths of $N$ and $\Delta$ resonances from the PFSM
calculation. In this table the theoretical masses of the baryon states 
as produced by the respective CQMs have been used. For the GBE CQM only 
two decay widths, namely $N$(1535) and $N$(1710), apparently coincide
with experimental data within their error bars. All the other ones are smaller
than experiment. In most cases a considerable underestimation of the
experimental data is found. The situation is similar for the OGE CQM, where
only the $N$(1710) coincides with experiment. Again, all other 
predictions remain (considerably) smaller than the data. The situation 
is even worse for the II CQM, calculated in the Bethe-Salpeter
approach~\cite{Metsch:2004qk},
where neither one of the predictions strictly agrees with experiment; 
each one is too small, some by far. In general, all relativistic 
calculations, independently of the framework applied, show similar
characteristics: They yield results always smaller than the 
experimental data or at most reaching their values from below.

\renewcommand{\arraystretch}{1.5}
\begin{table}
\caption{Covariant predictions for $\pi$ decay widths by the GBE
CQM~\cite{Glozman:1998ag} and the OGE CQM~\cite{Theussl:2000sj} along the
PFSM in comparison to experiment~\cite{Eidelman:2004wy} and a relativistic
calculation for the II CQM along the Bethe-Salpeter approach~\cite{Metsch:2004qk}.
In the last two columns the nonrelativistic results from an EEM are 
given. In all cases the theoretical resonance masses as predicted by the various 
CQMs have been used in the calculations.
\label{tab1}
}
\begin{ruledtabular}
\begin{tabular}{crccccc}
 Decay      &   Experiment   &  \multicolumn{3}{c}{Relativistic}
       &   \multicolumn{2}{c}{Nonrel. EEM } \\
{\small  $\rightarrow N\pi$}      &  {\small [MeV]}  & {\small GBE}  
          & {\small OGE} 
          & {\small II}
& {\small GBE} 
& {\small OGE}
\\
\hline
{\small $N(1440)$}
	& {  $\left(227\pm 18\right)_{-59}^{+70}$ }
             		 &  $33$
		          &  $68$
			 &  $38$
			 &  $6.7$
			 &  $27$\\
{\small $N(1520)$}
	 & { $\left(66\pm 6\right)_{-\phantom{0}5}^{+\phantom{0}9}$ }
			 &  $17$
			 &  $16$
			 &  $38$
			 &  $38$
			 &  $37$
\\
{\small $N(1535)$}
	 & { $ \left(67\pm 15\right)_{-17}^{+28}$ }
			 &  $90$
			 &  $119$
			 &  $33$
			 &  $554$
			 &  $1183$
\\
{\small $N(1650)$}
	 & { $ \left(109\pm 26 \right)_{-\phantom{0}3}^{+36}$ }
  			 &  $29$
			 &  $41$
			 &  $3$
			 &  $160$
			 &  $358$
\\
{\small $N(1675)$}
	 & { $ \left(68\pm 8\right)_{-\phantom{0}4}^{+14}$ }
			 &  $5.4$
			 &  $6.6$
			 &  $4$
			 &  $13$
			 &  $16$
\\
{\small $N(1700)$}
	 & { $ \left(10\pm 5\right)_{-\phantom{0}3}^{+\phantom{0}3}$ }
			 &  $0.8$
			 &  $1.2$
			 &  $0.1$
			 &  $2.2$
			 &  $2.7$
\\
{\small $N(1710)$}
	 & { $\left(15\pm 5\right)_{-\phantom{0}5}^{+30}$  }
			 &  $5.5$
			 &  $4.6$
			 &  $n/a$
			 &  $8.1$
			 &  $5.8$
\\
{\small $\Delta(1232)$}
	 & { $\left(119\pm 1 \right)_{-\phantom{0}5}^{+\phantom{0}5}$ }
			 &  $37$
			 &  $32$
			 &  $62$
			 &  $89$
			 &  $84$
\\
{\small $\Delta(1600)$}
	 & { $\left(61\pm 26\right)_{-10}^{+26} $ }
			 &  $0.1$
			 &  $1.8$
			 &  $n/a$
			 &  $92$
			 &  $85$
\\
{\small $\Delta(1620)$}
	 & { $\left(38\pm 8 \right)_{-\phantom{0}6}^{+\phantom{0}8}$ }
			 &  $11$
			 &  $15$
			 &  $4$
			 &  $77$
			 &  $178$
\\
{\small $\Delta(1700)$}
	 & { $ \left(45\pm 15\right)_{-10}^{+20}$ }
			 &  $2.3$
			 &  $2.3$
			 &  $2$
			 &  $11$
			 &  $9.2$
\\
\end{tabular}
\end{ruledtabular}
\end{table}

For a comparison to nonrelativistic results we advocate the elementary 
emission model (EEM), which can be viewed as the nonrelativistic 
analogue of the spectator model used for our covariant calculations. 
The comparison of the relativistic and nonrelativistic results in 
Table~\ref{tab1} tells us that there are huge differences between them. 
While there is a common trend in the relativistic results 
(practically no one of the predictions overshoots the data), the
nonrelativistic decay widths scatter below and above the experimental 
values. Incidentally, for the nonrelativistic results agreement with 
experiment is found in more cases than for the relativistic ones.
However, this observation should not be interpreted as a 
better quality of the nonrelativistic results. From the viewpoint of 
theory the relativistic calculations are much more appealing. In 
particular, the corresponding predictions are covariant. Furthermore, 
the fact that they generally underestimate the data may turn out as an 
advantage, especially when a more complete decay operator will be employed
than the simple spectator model used here. 

\renewcommand{\arraystretch}{1.5}
\begin{table}[b]
\caption{Same as Table~\ref{tab1} but with experimental resonance masses
instead of the theoretical ones.
\label{tab2}
}
\begin{ruledtabular}
\begin{tabular}{crcccc}
 Decay &   Experiment   &  \multicolumn{2}{c}{Relativistic}
       &   \multicolumn{2}{c}{Nonrel. EEM } \\
  {\small $\rightarrow N\pi$}    &  {\small [MeV]} & {\small GBE}  
          & {\small OGE} 
          & {\small GBE}
          & {\small OGE} 
\\
\hline
{\small $N(1440)$}
	& {  $\left(227\pm 18\right)_{-59}^{+70}$ }
	                    &  $30$
		          &  $37$
			 &  $6.2$
			 &  $14$
\\
{\small $N(1520)$}
	 & { $\left(66\pm 6\right)_{-\phantom{0}5}^{+\phantom{0}9}$ }
                             &  $17$
			 &  $16$
			 &  $38$
			 &  $36$
\\
{\small $N(1535)$}
	 & { $ \left(67\pm 15\right)_{-17}^{+28}$ }
                             &  $93$
			 &  $123$
			 &  $574$
			 &  $1230$
\\
{\small $N(1650)$}
	 & { $ \left(109\pm 26 \right)_{-\phantom{0}3}^{+36}$ }
                             &  $29$
			 &  $38$
			 &  $160$
			 &  $332$
\\
{\small $N(1675)$}
	 & { $ \left(68\pm 8\right)_{-\phantom{0}4}^{+14}$ }
                             &  $6.0$
			 &  $6.2$
			 &  $15$
			 &  $15$
\\
{\small $N(1700)$}
	 & { $ \left(10\pm 5\right)_{-\phantom{0}3}^{+\phantom{0}3}$ }
                             &  $0.9$
			 &  $1.2$
			 &  $2.9$
			 &  $2.9$
\\
{\small $N(1710)$}
	 & { $\left(15\pm 5\right)_{-\phantom{0}5}^{+30}$  }
                             &  $4.1$
			 &  $2.3$
			 &  $6.0$
			 &  $3.2$
\\
{\small $\Delta(1232)$}
	 & { $\left(119\pm 1 \right)_{-\phantom{0}5}^{+\phantom{0}5}$ }
                             &  $34$
			 &  $32$
			 &  $81$
			 &  $84$
\\
{\small $\Delta(1600)$}
	 & { $\left(61\pm 26\right)_{-10}^{+26} $ }
                             &  $0.1$
			 &  $0.5$
			 &  $56$
			 &  $30$
\\
{\small $\Delta(1620)$}
	 & { $\left(38\pm 8 \right)_{-\phantom{0}6}^{+\phantom{0}8}$ }
                             &  $10$
			 &  $15$
			 &  $75$
			 &  $178$
\\
{\small $\Delta(1700)$}
	 & { $ \left(45\pm 15\right)_{-10}^{+20}$ }
                             &  $2.9$
			 &  $3.1$
			 &  $14$
			 &  $15$
\\
\end{tabular}
\end{ruledtabular}
\end{table}

\renewcommand{\arraystretch}{1.5}
\begin{table*}[t]
\caption{
Covariant predictions for $\pi$ decay widths of the GBE, OGE, and
II CQMs (as in Table \ref{tab1}) presented as percentages of the experimental $\pi$
decay widths in comparison to experimental $N\pi\pi$ branching ratios.
\label{tab3}
}
\begin{ruledtabular}
{\begin{tabular}{@{}ccrccccccc@{}}
Decays&$J^P$&Experiment&\multicolumn{3}{c}{
Relativistic
}
&\multicolumn{3}{c}{\% of Exp. Width}&Experimental\\
{$\rightarrow N\pi $}&{}&  {\small [MeV]} &GBE & OGE & II & GBE & OGE & II 
& {\small $N\pi\pi$ Branching Ratio}\\
[0.25ex]
\hline
$N(1440)$
&
$\frac{1}{2}^+$&
$\left(227\pm 18\right)_{-59}^{+70}$ &
$33$ &
$68$ &
$38$ &
$14$ &
$30$ &
$17$ &
$30-40\%$
\\ 
$N(1520)$
&
$\frac{3}{2}^-$&
$\left(66\pm 6\right)_{-\phantom{0}5}^{+\phantom{0}9}$&
$17$ &
$16$ &
$38$ &
$26$ &
$24$ &
$58$ &
$40-50\%$  
\\ 
$N(1535)$
&
$\frac{1}{2}^-$&
$ \left(67\pm 15\right)_{-17}^{+28}$&
$90$ &
$119$ &
$33$ &
$134$ &
$178$ &
$49$   &
$1-10\%$  
\\ 
$N(1650)$
&
$\frac{1}{2}^-$&
$\left(109\pm 26 \right)_{-\phantom{0}3}^{+36}$&
$29$ &
$41$ &
$\phantom{0}3$ &
$27$ &
$38$ &
$\phantom{0}3$&
$10-20\%$ 
\\ 
$N(1675)$
&
$\frac{5}{2}^-$&
$ \left(68\pm 8\right)_{-\phantom{0}4}^{+14}$&
$5.4$ &
$6.6$ &
$\phantom{0}4$ &
$\phantom{0}8$ &
$10$ &
$\phantom{0}6$  &
$50-60\%$  
\\ 
$N(1700)$
&
$\frac{3}{2}^-$&
$ \left(10\pm 5\right)_{-\phantom{0}3}^{+\phantom{0}3}$&
$0.8$ &
$1.2$ &
$0.1$ &
$\phantom{0}8$ &
$12$ &
$\phantom{0}1$ &
$85-95\%$ 
\\ 
$N(1710)$
&
$\frac{1}{2}^+$&
$\left(15\pm 5\right)_{-\phantom{0}5}^{+30}$&
$5.5$ &
$4.6$ &
$n/a$ &
$37$ &
$31$ &
$n/a$ &
$40-90\%$  
\\ 
$\Delta(1232)$
&
$\frac{3}{2}^+$&
$\left(119\pm 1 \right)_{-\phantom{0}5}^{+\phantom{0}5}$&
$37$ &
$32$ &
$62$ &
$31$ &
$27$ &
$52$ &
$n/a$ 
\\ 
$\Delta(1600)$
&
$\frac{3}{2}^+$&
$\left(61\pm 26\right)_{-10}^{+26}$&
$0.07$ &
$1.8$ &
$n/a$ &
$\approx 0$ &
$3$ &
$n/a$ &
$75-90\%$  
\\ 
$\Delta(1620)$
&
$\frac{1}{2}^-$&
$\left(38\pm 8\right)_{-\phantom{0}6}^{+\phantom{0}8}$&
$11$ &
$15$ &
$\phantom{0}4$ &
$29$ &
$39$ &
$11$ &
$70-80\%$ 
\\ 
$\Delta(1700)$
&
$\frac{3}{2}^-$&
$\left(45\pm 15\right)_{-10}^{+20}$&
$2.3$ &
$2.3$ &
$\phantom{0}2$ &
$\phantom{0}5$ &
$\phantom{0}5$ &
$\phantom{0}4$ &
$80-90\%$ 
\\
\end{tabular}}
\end{ruledtabular}
\end{table*}

In the literature, results for decay widths are often calculated employing
phenomenological resonances masses instead of the theoretical ones (as
predicted by the respective CQM). Therefore, in
Table~\ref{tab2} we also give the decay widths calculated with the physical
resonance masses (but with the same CQM wave functions as before).
For the GBE CQM only slight variations are seen as
compared to Table~\ref{tab1}. This is not surprising, since the GBE CQM yields
a rather good reproduction of the experimental resonance masses. For the OGE
CQM some bigger deviations are observed, especially in case of the
positive-parity resonances $N$(1440), $N$(1710), and $\Delta$(1600),
where differences of more than 50\% may occur. For the OGE CQM the mass-shift
effect is also visible for the negative-parity resonance $\Delta$(1700).
Also the nonrelativistic results exhibit an analogous behaviour. We learn
that the resonance masses have a pronounced influence on the magnitudes of
the decay widths. For a given resonance (and a given wave function), the
decay width will come out bigger the larger its mass. It is therefore
an essential prerequisite that any CQM reproduces the
excitation spectra in fair agreement with experiment. 

By comparing the GBE and OGE columns in Table~\ref{tab2} one can see the
influences of different wave functions on the decay widths (since here
the employed resonance masses are the same in both cases, namely, the
experimental ones). Obviously the different components in the respective
wave functions also can have a respectable effect.
E.g., the OGE wave function produces a considerably larger $\pi$ decay width
for $N$(1535) than the GBE. A similar behaviour is found for $\Delta$(1620),
and to some extent also for $N$(1650). All of these
resonances have $J^P=\frac{1}{2}^-$. The OGE result is also higher in case of
the Roper resonances $N$(1440) and $\Delta$(1600). All the other $\pi$ decay
widths are of very similar magnitudes for both types of wave functions. Only,
in case of the $N$(1710) resonance the result with the GBE wave function
comes out appreciable larger than for the OGE. An analogous behaviour is found
in the nonrelativistic results for the EEM (with the only exception of
$\Delta$(1600)).  

It is interesting to examine the theoretical results from a different viewpoint.
In Table~\ref{tab3} we have presented the covariant predictions of the various
CQMs as percentage values relative to the experimental $\pi$ decay widths.
Evidently, since all of the predictions tend to be too small by their
absolute values, also these percentages turn out too small. The only
exception is $N$(1535). In this case an appreciable percentage is reached
(evidently because here the theoretical prediction is of the magnitude of
the experimental decay width). If we look at the corresponding experimental
$N\pi\pi$ branching ratio, incidentally, we observe that it is very small.
On the other hand, the $N\pi\pi$ branching ratios are observed to be quite
big in other cases, such as $N$(1675), $N$(1700), $\Delta$(1600), and
$\Delta$(1700). Here, they may become 60 to 90\%. Interestingly, in these
cases the theoretical $\pi$ decay widths assume only very small percentages
of the experimental decay widths. While the situation is not so clear-cut
with respect to the $N$(1440), $N$(1520), and $N$(1710) resonances -- they
appear to be intermediate in this behaviour -- one would have expected the
$N$(1650) decay width to be larger. Its $N\pi\pi$ branching ratio remains
smaller than 20\%. In this case, however, we should also observe that the
$N\eta$ decay width is of an appreciable magnitude experimentally and,
in addition, it results by far too large in the CQMs (see the discussion
below).

In the context of this comparison, looking at the $\pi$ decay widths relative
to the magnitudes of the branching ratios to other decay channels, we identify
a principal shortcoming of the present approach to mesonic decays.
A single-channel decay operator appears to be insufficient and a more complete
decay mechanism, including channel couplings, is called for.     

In Table~\ref{tab4} we also present the covariant predictions of the GBE and OGE
CQMs for $\eta$ decay widths. Again the theoretical resonance masses have been
employed, and a comparison to the nonrelativistic EEM is given. In the $\eta$
decay mode only two resonances, $N$(1535) and $N$(1650), show a sizable decay
width. This behaviour is exactly met by the CQMs. In particular, the $N$(1535)
decay width is reproduced within the experimental error bars by both relativistic
CQMs. One should recall that this is the same resonance for which also the
$\pi$ decay widths were reproduced best, practically in agreement with experiment
(see Table~\ref{tab1}). The experimental $\eta$ decay width of $N$(1650) is
overshooted by both CQMs. These deficiencies might be connected with the ones
in the $\pi$ decay widths, which came out unexpectedly small. Again large
differences are found between the covariant predictions and the nonrelativistic
EEM results.

\renewcommand{\arraystretch}{1.5}
\begin{table}[t!]
\caption{Covariant predictions for $\eta$ decay widths by the GBE
CQM~\cite{Glozman:1998ag} and the OGE CQM~\cite{Theussl:2000sj} along the
PFSM in comparison to experiment~\cite{Eidelman:2004wy}.
In the last two columns the nonrelativistic results from an EEM are 
given. In all cases the theoretical resonance masses as predicted by the various 
CQMs have been used in the calculations.
\label{tab4}
}
\begin{ruledtabular}
\begin{tabular}{crccccc}
 Decay &   Experiment   &  \multicolumn{2}{c}{Relativistic }
       & 
        \multicolumn{2}{c}{Nonrel. EEM } \\
  {\small $\rightarrow N\eta$}    &  {\small [MeV]} & {\small GBE}  
          & {\small OGE} 
          & {\small GBE}
          & {\small OGE} 
\\
\hline
{\small $N(1520)$}
	 & { $\left(0.28\pm 0.05\right)_{-0.01}^{+0.03}$ }
                             &  $0.04$
			 &  $0.04$
			 &  $0.04$
			 &  $0.04$
\\
{\small $N(1535)$}
	 & { $ \left(64\pm 19\right)_{-\phantom{0}28}^{+\phantom{0}28}$ }
                             &  $30$
			 &  $39$
			 &  $127$	
			 &  $236$		
\\
{\small $N(1650)$}
	 & { $ \left(10\pm 5 \right)_{-\phantom{00}1}^{+\phantom{00}4}$ }
                             &  $71$
			 &  $109$
			 &  $285$
			 &  $623$
\\
{\small $N(1675)$}
	 & { $ \left(0\pm 1.5\right)_{-\phantom{0}0.1}^{+\phantom{0}0.3}$ }
                             &  $0.6$
			 &  $0.9$
			 &  $1.1$
			 &  $1.8$
\\
{\small $N(1700)$}
	 & { $ \left(0\pm 1\right)_{-\phantom{0}0.5}^{+\phantom{0}0.5}$ }
                             &  $0.2$
			 &  $0.4$
			 &  $0.2$
			 &  $0.3$
\\
{\small $N(1710)$}
	 & { $\left(6\pm 1\right)_{-\phantom{00}4}^{+\phantom{0}11}$  }
                             &  $1.0$
			 &  $1.6$
			 &  $2.9$
			 &  $9.3$
\\
\end{tabular}
\end{ruledtabular}
\end{table}

Table~\ref{tab5} shows the $\eta$ decay widths calculated with experimental
resonance masses. By comparing to Table~\ref{tab4} the mass-shift effects are
clearly visible (in the same manner as for the $\pi$ decay widths above).
Again, the comparison of the GBE and OGE columns in Table~\ref{tab5} shows
the influences of the different CQM wave functions. For both the $N$(1535)
and the $N$(1650) the OGE wave function leads to higher values for the $\eta$
decay widths. In all other cases the predictions are very similar (and
small). A completely analogous behaviour is found for the nonrelativistic
EEM.

\renewcommand{\arraystretch}{1.5}
\begin{table}[h]
\caption{Same as Table~\ref{tab3} but with experimental resonance masses
instead of the theoretical ones.
\label{tab5}
}
\begin{ruledtabular}
\begin{tabular}{crcccc}
 Decay &   Experiment   &  \multicolumn{2}{c}{Relativistic}
       &  \multicolumn{2}{c}{Nonrel. EEM } \\
  {\small $\rightarrow N\eta$}    &  {\small [MeV]} & {\small GBE}  
          & {\small OGE} 
          & {\small GBE}
          & {\small OGE} 
\\
\hline
{\small $N(1520)$}
	 & { $\left(0.28\pm 0.05\right)_{-0.01}^{+0.03}$ }
                             &  $0.04$
			 &  $0.03$
			 &  $0.05$
			 &  $0.04$
\\
{\small $N(1535)$}
	 & { $ \left(64\pm 19\right)_{-\phantom{0}28}^{+\phantom{0}28}$ }
                             &  $36$
			 &  $46$
			 &  $155$
			 &  $282$
\\
{\small $N(1650)$}
	 & { $ \left(10\pm 5 \right)_{-\phantom{00}1}^{+\phantom{00}4}$ }
                             &  $72$
			 &  $95$
			 &  $288$
			 &  $543$
\\
{\small $N(1675)$}
	 & { $ \left(0\pm 1.5\right)_{-\phantom{0}0.1}^{+\phantom{0}0.3}$ }
                             &  $0.8$
			 &  $0.8$
			 &  $1.6$
			 &  $1.5$
\\
{\small $N(1700)$}
	 & { $ \left(0\pm 1\right)_{-\phantom{0}0.5}^{+\phantom{0}0.5}$ }
                             &  $0.4$
			 &  $0.4$
			 &  $0.4$
			 &  $0.3$
\\
{\small $N(1710)$}
	 & { $\left(6\pm 1\right)_{-\phantom{00}4}^{+\phantom{0}11}$  }
                             &  $1.0$
			 &  $1.4$
			 &  $2.2$
			 &  $4.6$
\\
\end{tabular}
\end{ruledtabular}
\end{table}

In this section we have presented covariant results for $\pi$ and $\eta$
decay widths as direct predictions by the relativistic GBE and OGE CQMs with the
PFSM decay operator. We have discussed them in comparison to experiment and
(as far as existing) to covariant results by the II CQM. Huge differences were
found as compared to nonrelativistic predictions along the EEM.

\section{Comparison to other decay calculations}
{\label{sec:comparison}}

Let us now have a view at our results in the light of mesonic decay calculations
existing in the literature. The EEM has been used in the comparisons of the
previous section as a (nonrelativistic) reference model, since it serves as
the best analogue of the relativistic PFSM. However, in the past one has also
learned that the EEM is not sufficiently sophisticated in order to provide a
reasonable description of the mesonic decays. A more elaborate decay mechanism
is furnished by the so-called pair-creation model (PCM). Corresponding studies
were performed, for example, by Stancu and Stassart~\cite{Stancu:1988gb},
by Capstick and Roberts~\cite{Capstick:1993th}, and by Theu{\ss}l, Wagenbrunn,
Desplanques, and Plessas~\cite{Theussl:2000sj}. In all of these calculations
some adjustments or additional parametrizations were applied on top of
the direct predictions by the CQMs employed. For instance, one introduced
by phenomenological parametrizations different forms and extensions of the
interaction (meson-creation) vertex or one adjusted the pair-creation 
strength. Mostly the size of the $\pi$ decay width of the $\Delta$(1232) was
used as a reference. In the works of SS and CR additional input was
advocated to fit further ingredients in the calculations that were
not determined by the underlying CQM.

Clearly, since the $\pi$ decay width of the $\Delta$(1232) was always used as
a constraint in the fits, this quantity is usually realistic in the PCM
calculations. We have therefore decided to scale the PFSM results in an
analogous manner by adjusting the quark-pion coupling in the decay operator
so as to reproduce the  $\Delta$(1232) in coincidence with experiment
(this corresponds to a value $g_{qqm}^2/{4 \pi}=2.15$ in case of the GBE CQM
and $g_{qqm}^2/{4 \pi}=2.49$ in case of the OGE CQM).
Table~\ref{tab6} gives the corresponding comparison of the results. Evidently,
all the PFSM decay widths are now scaled to larger values with the consequence
that the comparison to experiment is much improved. In particular, for the
GBE CQM the $\pi$ decay widths of $N$(1520), $N$(1650), $N$(1700), $N$(1710),
and $\Delta$(1620) now appear to be correct. One is therefore tempted to
accept that the tuning of the quark-pion coupling in the PFSM decay operator
improves the description of the decay widths. The PFSM calculation is now
at least of a similar overall quality in reproducing the data
as the PCMs. While there is no common trend in the PCM calculations by the
different groups, the scaled PFSM decay widths are either correct or still
remain smaller than the experimental data, with the notable exception of
$N$(1535). The $N$(1535) decay width that was correct before (see the
previous section) is now grossly overestimated. Consequently, by tuning
the quark-pion coupling strength one can influence the predictions for the
decay widths.

\renewcommand{\arraystretch}{1.5}
\begin{table}[h]
\caption{\label{tab6}
Scaled predictions for $\pi$ decay widths by the GBE 
CQM~\cite{Glozman:1998ag} and OGE CQM~\cite{Theussl:2000sj} along the
PFSM in comparison to results existing in the literature from calculations
along PCMs by Stancu and Stassart~\cite{Stancu:1988gb} (SS), by Capstick and 
Roberts~\cite{Capstick:1993th} (CR), and by Theu{\ss}l, Wagenbrunn, Desplanques,
and Plessas~\cite{Theussl:2000sj} (TWDP).
}
\begin{ruledtabular}
\begin{tabular}{crccccc}
Decay & Experiment~\cite{Eidelman:2004wy} & 
\multicolumn{2}{c}{\small PFSM} &
\multicolumn{3}{c}{\small PCM} \\
 {\small $\rightarrow N\pi$}     &  {\small [MeV]}  &
           {\small GBE} & {\small OGE} 
           & {\small SS} & {\small CR} &  {\small TWDP}  \\
\hline
{\small $N(1440)$}
	& { $\left(227\pm 18\right)_{-59}^{+70}$ }
		&  $106$
		&  $253$ 
	          &  $433$
                   &  $493$
                   &  $517$	
\\
{\small $N(1520)$}
	 & { $\left(66\pm 6\right)_{-\phantom{0}5}^{+\phantom{0}9}$  }
			 &  $55$
			 &  $60$
                             &  $71$
			 &  $100$
			 &  $131$
\\
{\small $N(1535)$}
	 & { $ \left(67\pm 15\right)_{-17}^{+28}$ }
			 &  $290$
			 &  $443$                            
                             &  $40$
			 &  $207$
			 &  $336$
\\
{\small $N(1650)$}
	 & { $ \left(109\pm 26 \right)_{-\phantom{0}3}^{+36}$ }
			 &  $93$
			 &  $153$                            
                             &  $5.3$
			 &  $115$
			 &  $53$
\\
{\small $N(1675)$}
	 & { $ \left(68\pm 8\right)_{-\phantom{0}4}^{+14}$ }
			 &  $17$
			 &  $25$                            
                             &  $31$
			 &  $33$
			 &  $34$
\\
{\small $N(1700)$}
	 & { $ \left(10\pm 5\right)_{-\phantom{0}3}^{+\phantom{0}3}$ }
			 &  $2.6$
			 &  $4.5$                             
                             &  $17$
			 &  $36$
			 &  $6$
\\
{\small $N(1710)$}
	 & { $\left(15\pm 5\right)_{-\phantom{0}5}^{+30}$  }
			 &  $18$
			 &  $17$                             
                             &  $3.2$
			 &  $12$
			 &  $54$
\\
{\small $\Delta(1232)$}
	 & { $\left(119\pm 1 \right)_{-\phantom{0}5}^{+\phantom{0}5}$ }
			 &  $119$
			 &  $119$                             
                             &  $115$
			 &  $104$
			 &  $120$
\\
{\small $\Delta(1600)$}
	 & { $\left(61\pm 26\right)_{-10}^{+26} $ }
			 &  $0.2$
			 &  $6.7$                             
                             &  $0.04$
			 &  $40$
			 &  $43$
\\
{\small $\Delta(1620)$}
	 & { $\left(38\pm 8 \right)_{-\phantom{0}6}^{+\phantom{0}8}$ }
			 &  $35$
			 &  $56$                             
                             &  $0.4$
			 &  $21$
			 &  $26$
\\
{\small $\Delta(1700)$}
	 & { $ \left(45\pm 15\right)_{-10}^{+20}$ }
			 &  $7.4$
			 &  $8.6$                            
                             &  $23$
			 &  $27$
			 &  $28$
\\
\end{tabular}
\end{ruledtabular}
\end{table}

We have also studied the dependence of the results on the size of the
quark-pion coupling constant in more detail. It is well known that the
quark-meson coupling can vary in a certain range dependent on the way
it is deduced from the experimentally measured nucleon-meson couplings
(which also have uncertainties).
In ref.~\cite{Glozman:1998fs} an allowed range of the quark-pion coupling
constant of $0.67~\lesssim g_{qq\pi}^2/{4 \pi}\lesssim 1.19$ was determined. In the
actual parametrization of the GBE CQM the value $g_{qq\pi}^2/{4 \pi}=0.67$
was employed; the same value was used for the results in the previous Section
for consistency reasons. If we now take the liberty of changing the strength
of the quark-pion coupling in the decay operator of Eq.~\ref{eq:hadrcurr},
we can scale it such that the decay widths are all increased from the results
in Table~\ref{tab1}. Following the principle that neither one of the decay
widths exceeds the experimental range, we can allow $g_{qq\pi}^2/{4 \pi}$ to
assume the value of 0.82. In this case the decay width of $N$(1535), which
results largest as compared to experiment, is still within the experimental
range (cf. Table~\ref{tab7}). Evidently, all other decay widths get increased
too and thus come closer to the experimental values. If we push the value of
$g_{qq\pi}^2/{4 \pi}$ to the highest allowed value of about 1.2, the results
in the last column of Table~\ref{tab7} are obtained. Here, only the decay
width of $N$(1535) is overshooted, while all the other ones are still
improved.   

\renewcommand{\arraystretch}{1.5}
\begin{table}[ht]
\caption{\label{tab7}
Predictions for $\pi$ decay widths by the GBE
CQM~\cite{Glozman:1998ag} and OGE CQM~\cite{Theussl:2000sj} along the
PFSM for different magnitudes of the quark-meson coupling constant $g_{qq\pi}$.
}
\begin{ruledtabular}
\begin{tabular}{cr|cc|cc|cc}
Decay & Experiment &
\multicolumn{2}{c|}{\small $\frac{g_{qq\pi}^2}{4\pi}$=0.67} &
\multicolumn{2}{c|}{\small $\frac{g_{qq\pi}^2}{4\pi}$=0.82} &
\multicolumn{2}{c}{\small $\frac{g_{qq\pi}^2}{4\pi} $=1.19}\\
  {\small $\rightarrow N\pi$}     &  &
            {\small GBE} & {\small OGE} &
            {\small GBE} & {\small OGE} &
            {\small GBE} & {\small OGE}
  \\
\hline
{\small $N(1440)$}
	& { $\left(227\pm 18\right)_{-59}^{+70}$ }
		&  $33$
		&  $68$
	          &  $40$
                    &  $83$
                    &  $64$	
                    &  $131$	
\\
{\small $N(1520)$}
	 & { $\left(66\pm 6\right)_{-\phantom{0}5}^{+\phantom{0}9}$  }
			 &  $17$
			 &  $16$
                              &  $21$
			 &  $20$
			 &  $33$
			 &  $31$
\\
{\small $N(1535)$}
	 & { $ \left(67\pm 15\right)_{-17}^{+28}$ }
			 &  $90$
			 &  $119$
                              &  $110$
			 &  $145$
			 &  $174$
			 &  $230$
\\
{\small $N(1650)$}
	 & { $ \left(109\pm 26 \right)_{-\phantom{0}3}^{+36}$ }
			 &  $29$
			 &  $41$
                              &  $35$
			 &  $50$
			 &  $56$
			 &  $79$
\\
{\small $N(1675)$}
	 & { $ \left(68\pm 8\right)_{-\phantom{0}4}^{+14}$ }
			 &  $5.4$
			 &  $6.6$
                              &  $6.6$
			 &  $8.1$
			 &  $10$
			 &  $13$
\\
{\small $N(1700)$}
	 & { $ \left(10\pm 5\right)_{-\phantom{0}3}^{+\phantom{0}3}$ }
			 &  $0.8$
			 &  $1.2$
                              &  $1.0$
			 &  $1.5$
			 &  $1.5$
			 &  $2.3$
\\
{\small $N(1710)$}
	 & { $\left(15\pm 5\right)_{-\phantom{0}5}^{+30}$  }
			 &  $5.5$
			 &  $4.6$
                              &  $6.7$
			 &  $5.6$
			 &  $11$
			 &  $8.9$
\\
{\small $\Delta(1232)$}
	 & { $\left(119\pm 1 \right)_{-\phantom{0}5}^{+\phantom{0}5}$ }
			 &  $37$
			 &  $32$
                              &  $45$
			 &  $39$
			 &  $71$
			 &  $62$
\\
{\small $\Delta(1600)$}
	 & { $\left(61\pm 26\right)_{-10}^{+26} $ }
			 &  $0.1$
			 &  $1.8$
                              &  $0.1$
			 &  $2.2$
			 &  $0.2$
			 &  $3.5$
\\
{\small $\Delta(1620)$}
	 & { $\left(38\pm 8 \right)_{-\phantom{0}6}^{+\phantom{0}8}$ }
			 &  $11$
			 &  $15$
                              &  $13$
			 &  $18$
			 &  $21$
			 &  $29$
\\
{\small $\Delta(1700)$}
	 & { $ \left(45\pm 15\right)_{-10}^{+20}$ }
			 &  $2.3$
			 &  $2.3$
                              &  $2.8$
			 &  $2.8$
			 &  $4.4$
			 &  $4.4$
\\
\end{tabular}
\end{ruledtabular}
\end{table}

Regarding the PCM calculations discussed above one should also bear in
mind that they are not covariant.
In view of the large relativistic effects found in the PFSM study one must
therefore take the corresponding results with
some doubt. We conclude that considerable efforts are still needed in
order to find the proper decay mechanism/operator. Of course, such studies
must be done in a fully relativistic framework.

\section{Summary}
We have presented covariant predictions for $\pi$ and $\eta$ decay widths
of $N$ and $\Delta$ resonances by two different types of relativistic CQMs.
The results have been obtained by calculating the transition matrix elements
of the PFSM decay operator directly with the wave functions of the respective
CQMs, and no additional parametrization has been introduced a-priori. It
has turned out as a general trend of the relativistic predictions that
the experimental data are usually underestimated. These findings are
congruent with the ones recently made in a Bethe-Salpeter
approach~\cite{Metsch:2004qk}. The reproduction of the experimental data
by the CQMs can be improved by an additional tuning of the quark-meson
coupling in the PFSM decay operator.

We have determined large relativistic effects in the decay widths. Thus
it appears mandatory to perform any (future) investigation of mesonic decays
in a relativistic framework. In this respect, Poincar\'e-invariant relativistic
quantum mechanics provides a viable approach to treating CQMs.

Upon a closer examination of the $\pi$ decay widths we have detected a certain
correlation of their magnitudes to the $N\pi\pi$ branching ratios. Whenever
the latter are large, the theoretical decay widths result by far too small.
We take this finding as a hint to a principal shortcoming of the applied
decay mechanism. Very probably a more elaborate decay operator including
channel couplings is needed. In this regard, point-form relativistic
quantum mechanics opens the way towards a covariant treatment of a
coupled-channel system.
Corresponding investigations have already been performed in the meson
sector~\cite{Krassnigg:2003gh,Krassnigg:2004sp}. It will be an ambitious
goal to construct a coupled-channel theory also for baryons.   

\begin{acknowledgments}
This work was supported by the Austrian Science 
Fund (Project P16945).
\end{acknowledgments}

\appendix*
\section{Details of the calculation}
{\label{sec:details}}
Here we explain several details relevant in the evaluation of the decay widths
from the matrix elements of the decay operator. The notation follows the one
of ref.~\cite{Yndurain:1996aa} utilizing relativistically invariant 
scalar products of states, spinors etc.

The velocity states defined in Eq.~(\ref{eq:velstates}) have the following
completeness relation
\begin{eqnarray}
{\mathbf 1}=\sum_{\mu_{1}\mu_{2}\mu_{3}}\int{
    \frac{d^{3}v}{v_{0}}
    \frac{d^{3}k_{2}}{2\omega_{2}}
    \frac{d^{3}k_{3}}{2\omega_{3}}
    \frac{\left(\omega_{1}+\omega_{2}+\omega_{3}\right)^{3}}
{2\omega_{1}}
    }
    \phantom{000}
    &&
    \nonumber\\
    \times \left|v;\vec{k}_1,\vec{k}_2,\vec{k}_3;\mu_1,\mu_2,\mu_3
\right\rangle
    \left\langle 
    v;\vec{k}_1,\vec{k}_2,\vec{k}_3;\mu_1,\mu_2,\mu_3\right| \, ,
&&
\phantom{000}
\end{eqnarray}
and the corresponding orthogonality relation reads
\begin{eqnarray}
  && \left\langle v;\vec{k}_1,\vec{k}_2,\vec{k}_3;\mu_1,\mu_2,\mu_3
    |v';\vec{k}'_1,\vec{k}'_2,\vec{k}'_3;\mu'_1,\mu'_2,\mu'_3\right
\rangle
\nonumber \\
&&   =\frac{2\omega_{1}2\omega_{2}2\omega_{3}}
    {\left(\omega_{1}+\omega_{2}+\omega_{3}\right)^{3}}
     \delta_{\mu_{1}\mu'_{1}}\delta_{\mu_{2}\mu'_{2}}\delta_{\mu_{3}
\mu'_{3}}
\nonumber\\
&&     \times
     v_{0}\delta^{3}\left(\vec{v}-\vec{v}'\right)
    \delta^{3}\left(\vec{k}_{2}-\vec{k}'_{2}\right)
     \delta^{3}\left(\vec{k}_{3}-\vec{k}'_{3}\right) \, .
\end{eqnarray}
For the actual calculation one needs the overlap matrix element
\begin{multline}
\left\langle p'_1,p'_2,p'_3;\sigma'_1,\sigma'_2,\sigma'_3
    |v;\vec{k}_1,\vec{k}_2,\vec{k}_3;\mu_1,\mu_2,\mu_3\right
\rangle
=\\
\prod^3_{i=1}{
D^{\frac{1}{2}}_{\sigma'_i \mu_i}
\left[ R_W \left( k_i,B\left( v\right)\right)\right]2p_i^0 \delta\left(\vec{p}_i-\vec{p}'_i\right)
}\, .
\end{multline}
The velocity-state representation of the baryon eigenstates of
Eq.~(\ref{eq:eigenstates}) then becomes
\begin{eqnarray}
    \left\langle v;\vec{k}_1,\vec{k}_2,\vec{k}_3;\mu_1,\mu_2,\mu_3
    |V,M,J,\Sigma\right\rangle=
\phantom{0000000}&& \nonumber\\ 
     \frac{\sqrt{2}}{M} v_{0}\delta^{3}\left(\vec{v}-\vec{V}\right)
  \sqrt{\frac{2\omega_{1}2\omega_{2}2\omega_{3}}
    {\left(\omega_{1}+\omega_{2}+\omega_{3}\right)^{3}}
    }
   \Psi_{MJ\Sigma}\left(
   \vec{k}_i;\mu_i
   \right).
   \phantom{00}
&&
\end{eqnarray}
This representation has the advantage of separating  
the motion of the system as a whole and the internal 
motion. The latter is described by the wave function
$\Psi_{MJ\Sigma}\left(\vec{k}_i;\mu_i\right)$, which is also the rest-frame
wave function. It depends on the individual spin projections $\mu_1$, $\mu_2$,
$\mu_3$ and on the individual momenta $\vec k_1$, $\vec k_2$, $\vec k_3$,
restricted by $\sum{\vec k_i}=0$; it is normalized as
\begin{eqnarray}
   && \delta_{MM'}\delta_{JJ'}\delta_{\Sigma\Sigma'}
    =
    \sum_{\mu_{1}\mu_{2}\mu_{3}}
    \int{d^{3}k_{2}d^{3}k_{3}}
\nonumber\\
&&
    \times
   \Psi^{\star}_{M'J'\Sigma'}\left( 
    \vec{k}_i;\mu_i
   \right)
   \Psi_{MJ\Sigma}\left(
   \vec{k}_i;\mu_i
   \right) \, .
\end{eqnarray}
These wave functions are obtained by solving the eigenvalue problem of the
interacting mass operator $\hat M$. 

In the practical calculation of the decays one adheres to a special frame of
reference. For convenience one chooses the 
rest frame of the decaying resonance with the momentum transfer in 
z-direction \footnote{We emphasize that the final results are independent of
the frame of reference since the PFSM calculation is manifestly covariant.}.
In the chosen reference frame the boosts to be applied in the transition
matrix element are given by
\begin{eqnarray}
    B\left(v_{in}\right)&=&{\mathbf 1}_4\\
    B\left(v_{f}\right)&=&
    \begin{pmatrix}
           \cosh \Delta & 0 & 0 & \sinh \Delta\\
	   0            & 1 & 0 & 0\\
	   0            & 0 & 1 & 0\\
	   \sinh \Delta & 0 & 0 & \cosh \Delta\\
           \end{pmatrix}
\end{eqnarray}
where
\begin{eqnarray}
   \sinh \Delta&=&\frac{Q}{M} \\
%
   \cosh \Delta&=&\sqrt{
   1+\frac{Q^{2}}{M^{2}}
   }
\end{eqnarray}
and $M$ is the mass of the nucleon.

With these boost transformations we can rewrite the spectator conditions
in Eq.~(\ref{eq:hadrcurr}) as
\begin{equation}
    \label{eq:spectcond}
    2p_{i}^{0}\delta\left({\vec p}_{i}-{\vec p}'_{i}\right)
    =2\omega_{i}^{0}\delta
     \left(
     B^{-1}\left(v_{f}\right)B\left(v_{in}\right){\vec k}_{i}
    -{\vec k}'_{i} 
      \right) \, .
\end{equation}
For the active quark one obtains
\begin{eqnarray}
\sum_{\sigma_1\sigma'_1} D_{\sigma'_1\mu'_1}^{\star \frac{1}{2}}
\left\{R_W\left[k'_1;B\left(v_{f}\right)\right]\right\}
{\bar u}\left(p'_1,\sigma'_1\right)
\gamma_5\gamma^\mu \lambda_m \nonumber \\
\times u\left(p_1,\sigma_1\right) D_{\sigma_1\mu_1}^{\frac{1}{2}}
\left\{R_W\left[k_1;B\left(v_{in}\right)\right]\right\} = \nonumber \\ [3mm]
\bar u\left({k'_{1}},\mu'_{1}\right)
    S\left[B^{-1}\left(v_{f}\right)\right]
    \gamma_{5} \gamma^{\mu}{ \lambda_m}
    S\left[B\left(v_{in}\right)\right]
    u\left({ k_{1}},\mu_{1}\right) = \nonumber \\
\bar u\left({k'_{1}},\mu'_{1}\right)
    \left(\cosh\frac{\Delta}{2}-\gamma_{0}\gamma_{3}\sinh\frac{\Delta}{2}\right)
    \gamma_{5} \gamma^{\mu}{ \lambda_m}
    u\left({ k_{1}},\mu_{1}\right) \, , \nonumber \\
\end{eqnarray}
where the boost transformations on the Dirac spinors, represented by
$S\left[B^{-1}\left(v_{f}\right)\right]$ and $S\left[B\left(v_{in}\right)\right]$,
respectively, have been written out explicitly in the last line.
These expressions are to be used in the evaluation of the matrix element of
the decay operator, where the quark spinors can be represented conveniently
in the form 
\begin{equation}
    u\left({ k_{1}},\mu_{1}\right)=\sqrt{\omega_{1}+m_q}
    \begin{pmatrix}
           1\\
	   \frac{\sigma_{1} {\vec k_{1}}}{\omega_{1}+m_q}\\
           \end{pmatrix}
	   \chi\left(\mu_{1}\right) \, ,
\end{equation}
with $m_q$ the quark mass and the Pauli spinors 
$\chi\left(\mu_{1}\right)$ normalized to unity. 
%
%

%
\end{document}